\documentclass[a4paper,10pt,twoside]{cpc-hepnp}

\usepackage{multicol}
\usepackage{graphicx}
\usepackage{booktabs}
\usepackage{amssymb,bm,mathrsfs,bbm,amscd}
\usepackage[tbtags]{amsmath}
\usepackage{lastpage}
\usepackage{CJK}
\usepackage[usenames]{color}

\begin{document}
\begin{CJK*}{GBK}{song}

\fancyhead[c]{\small Chinese Physics C~~~Vol. 37, No. 1 (2013)
010201} \fancyfoot[C]{\small 010201-\thepage}

\footnotetext[0]{Received 14 March 2009}

\title{Tensor force impact on shell evolution in neutron-rich Si and Ni isotopes \thanks{The study was supported by the Interdisciplinary Scientific and Educational School of Moscow University "Fundamental and Applied Space Research". Sidorov S.V. expresses his gratitude for the support provided by the Foundation for the Development of Theoretical Physics and Mathematics "BASIS".}}

\author{%
S.V. Sidorov$^{1,2;1)}$\email{sv.sidorov@physics.msu.ru}%
\quad A.S. Kornilova$^{1}$
\quad T.Yu. Tretyakova$^{1,2}$
}
\maketitle

\address{%
$^1$ Faculty of Physics, Lomonosov Moscow State University, Moscow 119991, Russia\\
$^2$ Skobeltsyn Institute of Nuclear Physics, Lomonosov Moscow State University, Moscow 119991, Russia\\
}

\begin{abstract}
The influence of the tensor interaction of nucleons on the characteristics of neutron-rich silicon and nickel isotopes was studied in this work. Tensor forces are taken into account within the framework of the Hartree-Fock approach with the Skyrme interaction. It is shown that the addition of tensor component of interaction improves the description of the splitting between different single-particle states and leads to a decrease in nucleon-nucleon pairing correlations in silicon and nickel nuclei. Special attention was given to the role of isovector tensor forces relevant for interaction of like nucleons.
\end{abstract}



\footnotetext[0]{\hspace*{-3mm}\raisebox{0.3ex}{$\scriptstyle\copyright$}2013
Chinese Physical Society and the Institute of High Energy Physics
of the Chinese Academy of Sciences and the Institute
of Modern Physics of the Chinese Academy of Sciences and IOP Publishing Ltd}%

\begin{multicols}{2}

\section{Introduction}

The development of experimental methods with radioactive beams made it possible to significantly expand the isotope map in the regions of neutron or proton excess. When moving into the region of exotic nuclei, new and often unexpected phenomena were uncovered, such as the disappearance of the "classical" magic numbers $N=20$ and 28 and the appearance of new $N=14,$ 16, 32 or 34. Such profound changes in nuclear structure allow for further improvement and selection of theoretical models, motivating more studies of nuclear structure evolution using sufficiently complete chains of isotopes or isotones as an example. New problems arose in description of changes in the splittings between certain single-particle states in the framework of the usual mean-field theory using effective nucleon-nucleon ($NN$) forces. These issues compelled us to pay closer attention to the subtle features of the interaction. Thus, in recent decades, the role of the tensor $NN$ interaction in formation of some features in the structure of exotic nuclei has been actively studied \cite{Otsuka20}.

The most general form of nucleon-nucleon forces involves the contribution of the tensor component contained in all meson-exchange models. One of the most important experimental evidences for the presence of off-center forces is the non-zero quadrupole moment of the deuteron. Tensor interaction leads to an additional long-range spatial correlation of the wave functions of two nucleons in the triplet state and actually determines the existence of the deuteron \cite{BW52}: the state in which the vector connecting the neutron with the proton is aligned with their spins is the most energetically favorable. A simple account of the isospin dependence of nuclear forces shows that in the case of an isovector state, the attraction arising due to tensor forces should be three times weaker than in the state with zero isospin \cite{Otsuka20}.

Despite the crucial role of the tensor component in the structure of the deuteron, various models of many-particle systems based on effective interactions, such as Skyrme or Gogny forces, have not explicitly considered the contribution of the tensor interaction for a long time. On the one hand, the introduction of additional parameters complicates the computational procedure; on the other, fitting the interaction parameters to the experimental data of a large number of nuclei makes it possible to effectively take into account the influence of certain interaction features. However, following the accumulation of new data in experiments with radioactive beams, the question of the influence of the tensor contribution on the calculations of the nuclear structure in many-particle models has again become relevant.

The role of tensor forces in the formation of new magic numbers and its influence on the shape of various nuclei was investigated in a significant number of papers \cite{Ar60,St77,Colo07,Tar08,Sat09,Zal09}. It was shown in \cite{Otsuka20,Otsuka05} that the tensor neutron-proton interaction in the nucleus leads to attraction when the nucleon spins are parallel, and to repulsion in the opposite case. This property affects the spin-orbit splitting and changes the energies of single-particle states. In a number of cases, tensor forces were shown to play a crucial role in formation of the islands of inversion \cite{No21,Sm10}. Interplay between tensor interaction and pairing correlation was also reported to cause changes in the single-particle structure and nucleon density distribution with possible formation or disappearance of bubble structure in certain nuclei \cite{Nak13}. Changes in nuclear structure are also of importance in various applications in astrophysics. Examples include the impact on the strength distribution of Gamow-Teller resonances affecting the weak processes happening alongside the r-process in supernovae \cite{Dzh20}. A detailed analysis of the role of tensor interaction in weak processes taking place in stars undergoing a gravitational collapse is also given in \cite{Ha21,Ya23}.

It is important to note that it is standard to describe the neutron-proton interaction when discussing the role of tensor forces, namely: the influence of tensor forces on the position of single-particle levels of nucleons of one type is considered when nucleons of another type are added to the nucleus \cite{Otsuka20}. At the same time, significantly less attention was paid to the role of the isovector tensor component, which is also responsible for the interaction of identical nucleons. Some few works on this topic include \cite{Chen21}, where the authors study the impact of tensor forces on the shell structure and deformation of zirconium isotopes.

In this paper, we discuss the influence of tensor forces on various characteristics of even $^{28-42}$Si silicon isotopes and $^{50-78}$Ni nickel isotopes. All calculations are carried out in the self-consistent Skyrme-Hartree-Fock approach with forces that include the tensor interaction component. Previously, in \cite{Tar08}, the influence of tensor forces on the splitting of proton states in silicon isotopes was considered using the SLy5 parametrization in comparison with the approach of the relativistic mean field theory. The impact of tensor forces on deformation of silicon nuclei was also discussed in \cite{Li13}. Nickel isotopes with magic number $Z=28$ were studied more extensively in a variety of approaches (see, for example, \cite{Les07,Ban22,Ro21,Ot06}). Our task is to compare the results for different variants of $NN$ interaction, as well as to get an estimate of contribution of tensor forces to the splitting of not only proton, but also neutron single-particle levels, with increasing neutron excess. We also check the interplay between the effect of pairing of identical nucleons and the influence of tensor forces. The use of the Bardeen-Cooper-Schrieffer approach to describe pairing makes it possible to isolate the role of pairing effects and evaluate the change in the effect when tensor forces are turned on.

The article is structured as follows. In section 2, we review the basics of the Skyrme-Hartree-Fock approach and account for tensor forces in the approximation of zero-range forces. In section 3,
we discuss the impact tensor interaction has on various nuclei in the chains silicon and nickel isotopes. On the example of two interactions, SLy5 and SGII with and without forces, we demonstrate that the shell structure evolution in these isotopes can be very sensitive to both the central and tensor parts of the nucleon-nucleon interaction. Several other interactions were also employed in order to verify the effects tensor forces have on the splitting between various neutron and protons states in select isotopes.

\section{Model approach}

All calculations were performed within the framework of the Hartree-Fock approach with phenomenological interaction in the form of Skyrme forces. Pairing correlations were treated within the BCS model. The standard nucleon-nucleon Skyrme potential reads \cite{VB}:
\begin{align}
	V_{12} &= t_0 (1+x_0 P_\sigma) \delta(\vec{r}_1 - \vec{r}_2) \nonumber \\
	&+ \frac{1}{2} t_1 (1+x_1 P_\sigma) \left[ \vec{k}'^2 \delta(\vec{r}_1 - \vec{r}_2) + \delta(\vec{r}_1 - \vec{r}_2) \vec{k}^2 \right] \nonumber \\
	&+ t_2 (1+x_2 P_\sigma) \vec{k}' \delta(\vec{r}_1 - \vec{r}_2) \vec{k} \nonumber \\
	&+ \frac{1}{6} t_3 (1+x_3 P_\sigma) [\rho(\frac{1}{2}(\vec{r}_1 + \vec{r}_2))]^\gamma \delta(\vec{r}_1 - \vec{r}_2) \nonumber\\
	&+iW_0 \vec{\sigma} [\vec{k}' \delta(\vec{r}_1 - \vec{r}_2) \vec{k}].
	\label{Skyrme}
\end{align}
Here, $\vec{r}_1$ and $\vec{r}_2$ are nucleon coordinates,  $\vec{k}=\frac{1}{2i}(\vec{\nabla}_1-\vec{\nabla}_2)$, $\vec{k}'$ is the operator complex-conjugate to $\vec{k}$ (the corresponding conjugate operators $\vec{\nabla}'$ act on the wave function to the left), $\vec{\sigma}=\vec{\sigma}_1+\vec{\sigma}_2$, $P_\sigma = \frac{1}{2}(1+\vec{\sigma}_1\vec{\sigma}_2)$, $t_0..t_3$, $x_0..x_3$, $\gamma$ and $W_0$ are the interaction parameters.

The expression for the tensor nucleon-nucleon interaction of zero radius was proposed in the first works of Skyrme \cite{Skyr56}. It is of the form:
\begin{align}
	V_{tens} &= \frac{1}{2} t_e \{ [3(\sigma_1 \cdot \vec{k}')(\sigma_2 \cdot \vec{k}') - (\sigma_1 \cdot \sigma_2) \vec{k}'^2] \delta(\vec{r}_1 - \vec{r}_2) \nonumber \\
	&+ \delta(\vec{r}_1 - \vec{r}_2) [3(\sigma_1 \cdot \vec{k})(\sigma_2 \cdot \vec{k}) - (\sigma_1 \cdot \sigma_2) \vec{k}^2] \} \nonumber \\
	&+ t_o [3(\sigma_1 \cdot \vec{k}') \delta(\vec{r}_1 - \vec{r}_2) (\sigma_2 \cdot \vec{k}) - (\sigma_1 \cdot \sigma_2) \vec{k}' \delta(\vec{r}_1 - \vec{r}_2) \vec{k}],
	\label{tensor}
\end{align}
where $t_e$ and $t_o$ are tensor interaction parameters. The indices $e$ and $o$ correspond to the parity of states: the term $\sim t_e$ affects the states of a pair of nucleons with relative orbital momentum $L=0$ and $L=2$ ($S$ and $D$ wave), while the term $\sim t_o$ affects the states with $L=1$ and $L=3$ ($P$ and $F$ wave). Since the nucleon-nucleon tensor interaction works only in the state of a pair of nucleons with a total spin $S=1$, the parts $\sim t_e$ and $\sim t_o$ describe tensor effects in the isoscalar and isovector channels, respectively.

Relations \eqref{Skyrme} and \eqref{tensor} are used to obtain an expression for the energy density functional and, subsequently, the Hartree-Fock equations. The expression for the energy density corresponding to the nucleon-nucleon potential \eqref{Skyrme} is given in \cite{VB,Chab98}.

The contribution of tensor forces to the energy density comes in form of the so-called $J^2$-terms \cite{St77, Per04}:
\begin{align}
	\mathcal{H}^t = \frac{1}{2} \alpha (\vec{J}^2_n+\vec{J}^2_p) + \beta \vec{J}_n \cdot \vec{J}_p,
	\label{tens_en}
\end{align}
where $\vec{J}_{p,n}$ are proton and neutron spin densities:
\begin{align}
	\nonumber
	\vec{J}_q(\vec{r}) = (-i) \sum_{i, m_s, m_s'} \phi_i^*(\vec{r},m_s,q) [\nabla \times \vec{\sigma}] \phi_i(\vec{r},m_s',q).
\end{align}
Here, $\phi_i$ are single-particle wave functions, $m_s$ is the nucleon spin projection. Notably, $J^2$-terms arise from the exchange terms even without the account for tensor forces. Inclusion of tensor interaction results in the dependence of $\alpha, \beta$ on both the features of the central and tensor parts of the interaction \cite{Les07, Colo07}:
\begin{align}\nonumber
	\alpha &= \alpha_C + \alpha_T, \;\; \beta = \beta_C + \beta_T, \\\nonumber
	\alpha_C &= \frac{1}{8}(t_1-t_2) - \frac{1}{8}(t_1x_1 + t_2x_2), \\ \nonumber
	\beta_C &= - \frac{1}{8}(t_1x_1 + t_2x_2), \\ \nonumber
	\alpha_T &= \frac{5}{4} t_o, \\ \nonumber
	\beta_T &= \frac{5}{8} (t_e+t_o). 
\end{align}
Indeed, only the isovector part of the tensor force is relevant for pairs of like nucleons, while both the isovector and isoscalar components contribute to $np$-interaction.

The impact of tensor interaction on single-particle energies (SPEs) is considered in detail in the works of Otsuka \textit{et al}, \cite{Otsuka05,Otsuka20}. It was shown that if the neutron level $j'$ is filled in the given isotope chain, then the matrix elements of the tensor interaction $V^T_{j,j'}$ between the $j'$ neutrons and protons in $j_>=l+1/2$ and $j_ <=l-1/2$ states fulfill the following relation:
\begin{align}
	(2j_>+1) V^T_{j_> ,j'} + (2j_<+1) V^T_{j_< ,j'} = 0,
	\label{Otsuka}
\end{align}
where $T$ is the isospin of the nucleon pair. Furthermore, the filling of $j'_<$ with neutrons leads to an increase in the spin-orbit splitting between proton levels, and when $j'_>$ is filled, this splitting, on the contrary, decreases. Commonly used to describe the tensor part of the $np$ interaction, Otsuka's rule \eqref{Otsuka} is also valid for identical nucleons, although as of present there is still an on-going debate regarding the sign of the tensor force contributions in the isovector channel, and as such this is one of the questions we study extensively in this paper.

The effect of tensor forces on the spin-orbit splitting is due to the fact that their contribution to the one-particle potential is actually of similar structure to the contribution from the spin-orbit interaction. The sum of these two contributions, calculated as the first derivative of the energy density with respect to the nucleon density $\rho$, for protons (neutrons) reads \cite{Les07}:
\begin{align}
	W_{p(n)}(r) = \frac{W_0}{2} (2 \nabla \rho_{p(n)} + \nabla \rho_{n(p)}) + \alpha J_{p(n)} + \beta J_{n(p)}. 
\end{align}

In the $1d2s$-shell nuclei, pairing of identical nucleons plays an important role. In our calculations, we used the Bardeen-Cooper-Schrieffer (BCS) \cite{Suhonen} method, and the joint HF + BCS procedure was carried out in several iterations, each iteration including the solution of the self-consistent HF problem with subsequent application of the BCS scheme. A simple potential in the form of $\delta$-forces was taken as the pair correlation potential. The magnitude of the pair forces for each nucleus was chosen so that the energy gap obtained during the BCS procedure for a given even nucleus was equal to
\begin{align}
	\Delta_q = - \frac{1}{4} (S_q(A + 1)- 2S_q(A) + S_q(A - 1)),
\end{align}
where $S_q$ is the proton ($q=p$) or neutron ($q=n$) separation energy.

It should also be pointed out that we used the approximation of spherical symmetry. Experimental data indicate the presence of deformation in stable silicon isotopes, the value of the quadrupole deformation parameter $^{28}$Si is $\beta = -0.42\pm 0.02$ \cite{Hao84}. For other silicon isotopes, there is no experimental information other than estimates based on the strength of E2 transitions $B(E2)$. The same can be said about most isotopes of nickel, although data on $B(E2)$ points to the values of $\beta \simeq 0.1 \div 0.2$. In such a situation, the spherical approximation is a reasonable approach for model estimates and investigation of such features of nucleon interactions as tensor forces or nucleon pairing effects.

\section{Results}

Calculations of the single-particle structure of even silicon isotopes $^{28-42}$Si and nickel isotopes $^{56-78}$Ni were carried out using the SLy5+T \cite{Colo07} and SGII+T \cite{Sag09} parametrizations. It is important to note that the initial selection of these sets of parameters was conducted without taking into account the tensor interaction component. The SLy5 parametrization \cite{Chab98} was chosen to realistically describe the main characteristics of symmetric nuclear matter, as well as the binding energies and root-mean-square radii of doubly magic nuclei from oxygen to lead. The SGII interaction \cite{Sag81} was developed for the purposes of improved description of collective nuclear excitations. The parameters of the tensor interaction were obtained later in both cases, while the values of the initial parameters of the central part of the interaction were retained. Such a tactic, strictly speaking, violates the consistency of the parameter procedure, but, from the point of view of analyzing the effect of tensor forces on the structure of atomic nuclei, this kind of parametrization is more convenient, allowing comparison of the results of calculations with and without the tensor component. Notably, most currently existing parametrizations of Skyrme forces involving a tensor component agree on the signs of parameters $\alpha_T<0$ and $\beta_T>0$, and typically have their absolute value ranging up to around 200 MeV$\cdot$fm$^5$. In order to get a better understanding of how the $np$ and isovector tensor forces compare with each other, we carried out some additional calculations with several interactions, namely SLy4(+T), SAMi(+T) and SGII with different tensor forces (which we will denote as SGII+T2 in the remainder of the article to avoid confusion), exhibiting $\alpha_T$ and $\beta_T$ in this wide range.

\end{multicols}
\begin{center}
\tabcaption{ \label{tab1} Characteristics of nuclear matter for SLy5, SGII, SLy4 and SAMi parametrizations: saturation density $\rho_0$ (fm$^{-3}$), energy per nucleon $E_0$ (MeV), incompressibility $K_{\infty}$ (MeV) and symmetry energy $a_s$ (MeV), as well as the parameters of the central $\alpha_C$, $\beta_C$ and tensor $\alpha_T$, $\beta_T$ contributions in $J^2$-terms (MeV fm$^5$).}
\footnotesize
\begin{tabular*}{130mm}{@{\extracolsep{\fill}}cccccccccc}
	\toprule Interaction & Ref. & \,\, $\rho_0$ \,\, & \,\, $E_0$ \,\, & \,\, $K_{\infty}$  \,\, & \,\, $a_s $ \,\, &  \,\,  $\alpha_C $ \,\,  &  \,\, $\beta_C$ \,\,  & \,\,  $\alpha_T $ \,\, & \,\, $\beta_T$ \,\, \\
	\hline
	SGII & \cite{Sag81} & $0.158$ & $-15.60$ & $214.65$ & $26.83$ & $0$ & $0$ & $0$ & $0$ \\
	SGII+T & \cite{Sag09} & $0.158$ & $-15.60$ & $214.65$ & $26.83$ & $-5.434$ & $-53.171$ & $-180$ & $120$ \\
	SGII+T2 & \cite{Wu20} & $0.158$ & $-15.60$ & $214.65$ & $26.83$ & $-5.434$ & $-53.171$ & $-162.5$ & $4.17$ \\
	\hline
	SLy5 & \cite{Chab98} & $0.161$ & $-15.98$ & $229.92$ & $32.01$ & $80.2$ & $-48.9$ & $0$ & $0$ \\
	SLy5+T & \cite{Colo07} & $0.161$ & $-15.98$ & $229.92$ & $32.01$ & $80.2$ & $-48.9$ & $-170$ & $100$ \\
	\hline
	SLy4 & \cite{Chab98} & $0.16$ & $-15.97$ & $229.9$ & $32$ & $0$ & $0$ & $0$ & $0$ \\
	SLy4+T & \cite{Zal08} & $0.16$ & $-15.97$ & $229.9$ & $32$ & $81.79$ & $-47.37$ & $-105$ & $15$ \\
	\hline
	SAMi & \cite{Roc12} & $0.159$ & $-15.93$ & $245$ & $28$ & $101.88$ & $31.78$ & $0$ & $0$ \\
	SAMi+T & \cite{She19} & $0.164$ & $-16.15$ & $244$ & $29.7$ & $112.79$ & $35.13$ & $-39.80$ & $66.65$\\
	\bottomrule
\end{tabular*}%
\end{center}
\begin{multicols}{2}

Some characteristics of nuclear matter, as well as the parameters of tensor forces for the mentioned interactions, are given in Table~\ref{tab1}. The SGII+T parametrization serves as our choice of interaction with the largest contribution of tensor forces. Despite the fact that this parameter set does not always describe well the characteristics of nuclei, it is very interesting as a test variant. It is also important to note that, in addition to tensor contributions, there are also differences for the main parts of the parameter sets: the SLy family forces give more realistic values of the symmetry energy and incompressibility of nuclear matter.

\end{multicols}
\begin{center}
\includegraphics[width=8cm]{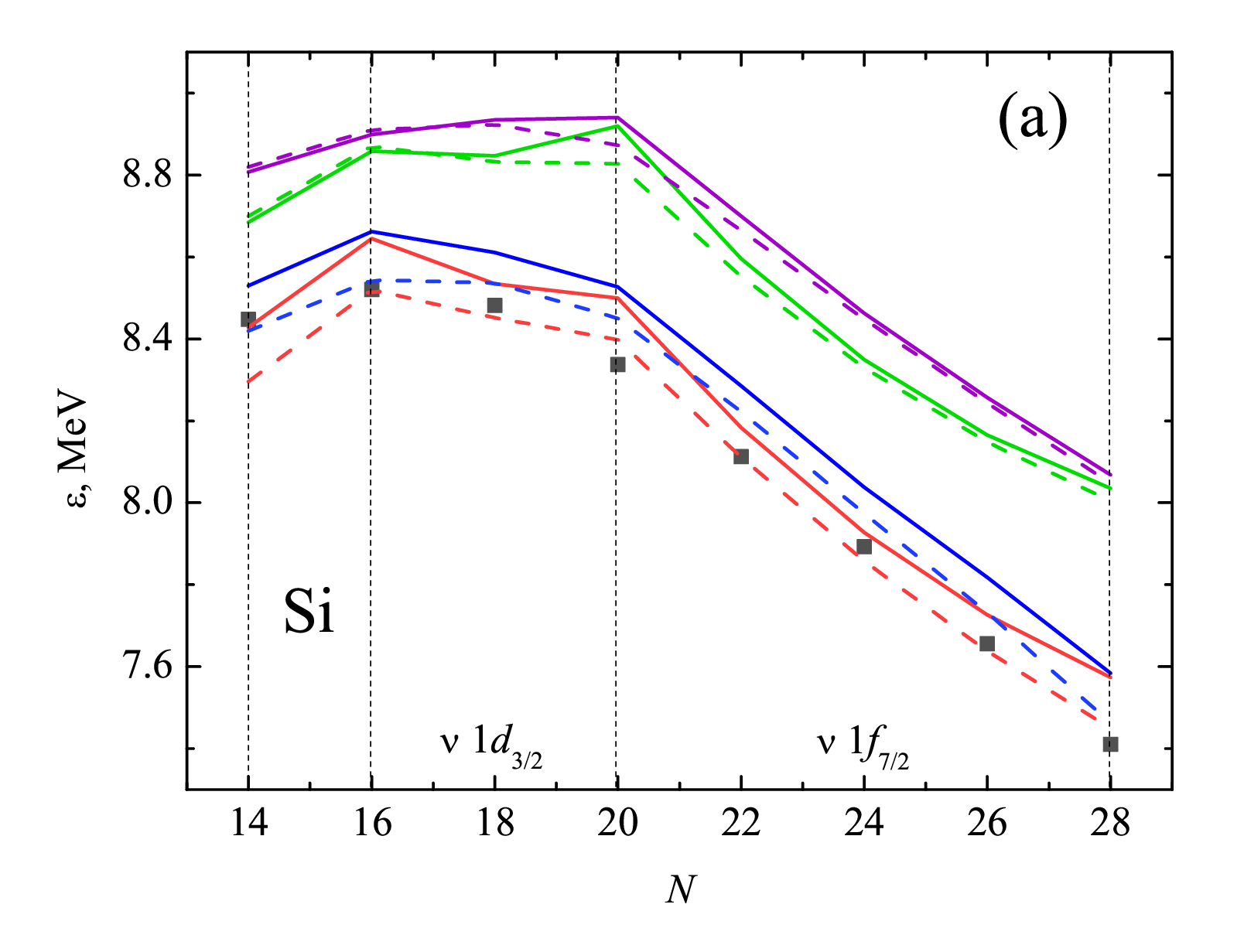}
\includegraphics[width=8cm]{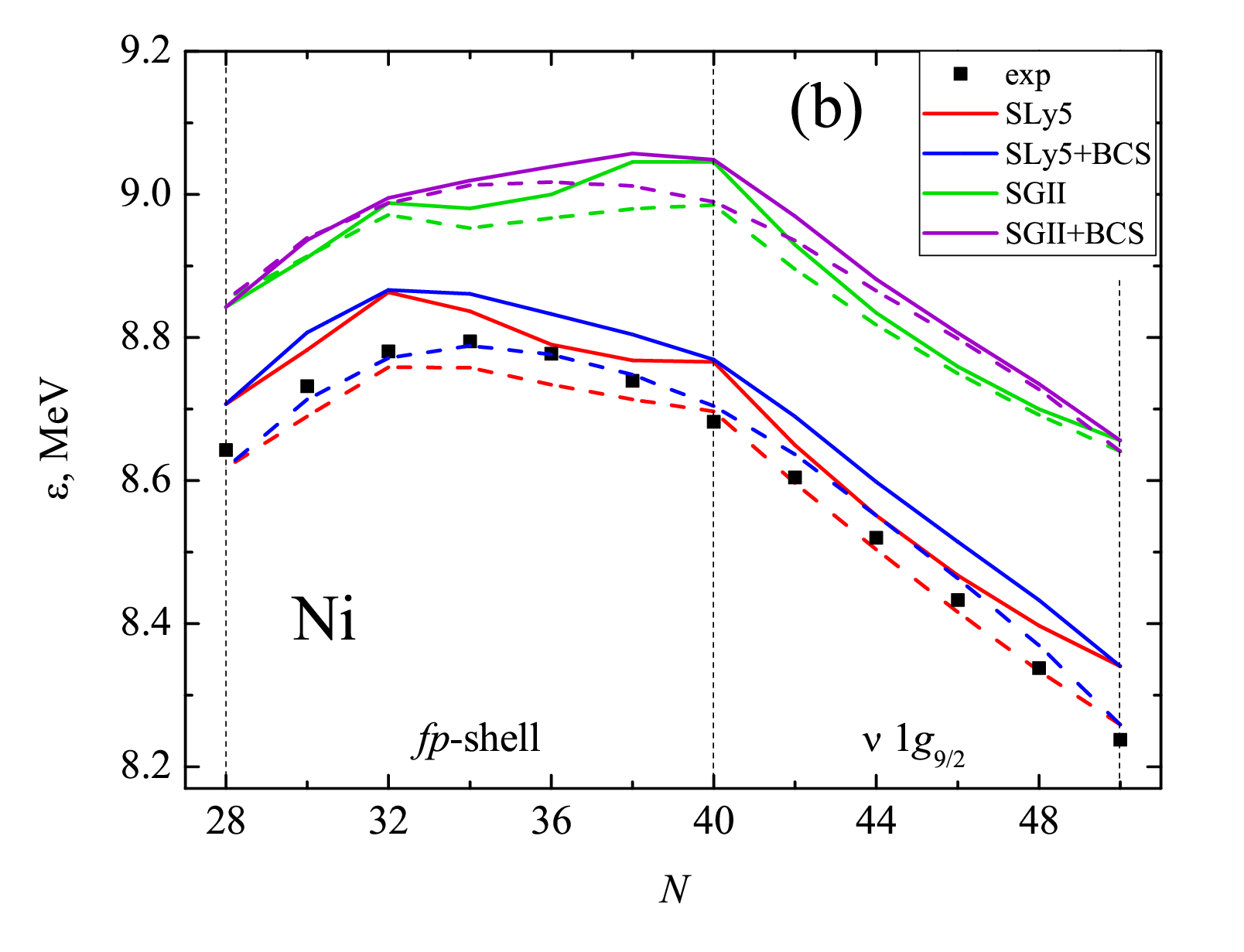}
\figcaption{\label{Si_BE} Binding energy per nucleon $\epsilon$ in even silicon (a) and nickel (b) isotopes with $N$ neutrons, with and without taking into account pairing correlations. Solid (dashed) lines show calculations with tensor forces (without taking tensor forces into account). Experimental data \cite{AME20} are marked with dots. (Colour online).}
\end{center}
\begin{multicols}{2}

The general scale of effects associated with taking tensor forces or pair correlations into account in comparison with the differences due to the properties of the central parts of the interaction, can be illustrated by the binding energy per nucleon $\varepsilon = B/A$ obtained with various $NN$-interactions (Fig.~\ref{Si_BE}). We see that the parametrization SLy5 is in the best agreement with the experimental values of the binding energies of silicon isotopes (Fig.~\ref{Si_BE}a) in a fairly wide range, while SGII overestimates the binding energy in all of the isotopes under consideration. The same can be said of the results for nickel isotopes (Fig.~\ref{Si_BE}b). Accounting for pairing, as well as the introduction of a tensor component, leads to overestimated values for almost all isotopes as well. Notably, introduction of pairing appears to smooth out the shell effects. Finally, it can be seen that the calculation results depend much more strongly on the choice of the central interaction, the effects of tensor forces and nucleon pairing are comparable in magnitude and have a smaller impact on the properties of the ground states of the considered nuclides.

\end{multicols}
\begin{center}
\includegraphics[width=16cm]{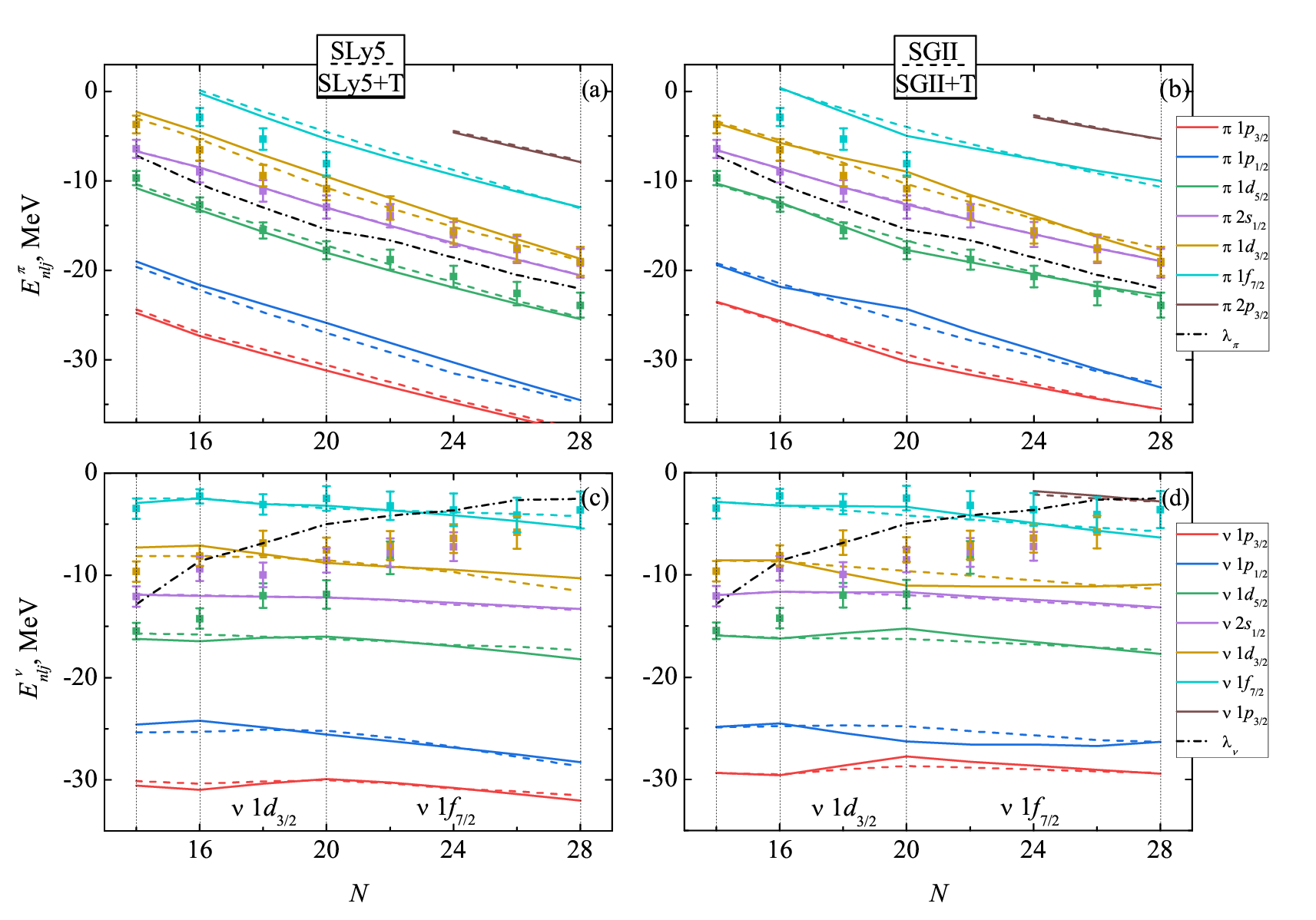}
\figcaption{\label{Si_spe} SPEs in even silicon isotopes: proton (a, b) and neutron (c, d) states. Solid (dashed) lines show calculations with tensor forces (without taking tensor forces into account). Experimental evaluated data \cite{Bes17} are marked with dots. The dashed-dotted line shows the chemical potential of protons (neutrons) in figures a and b (c and d).}
\end{center}
\begin{multicols}{2}

Fig.~\ref{Si_spe} shows the dependences of proton $E^\pi_{nlj}$ and neutron $E^\nu_{nlj}$ SPEs ($nlj$ denoting the quantum numbers) on the number of neutrons $N$ in even silicon isotopes $^{28-42}$Si obtained with the SLy5+T and SGII+T parametrizations. For comparison, the results of calculations are also given without taking into account the tensor contribution. Here and onwards, all of the shown results were obtained with pairing taken into account. Additionally, we show the value of the chemical potential of nucleons of the corresponding type that was estimated using the formula:
\begin{align}
	\lambda^{\textrm{(exp)}}_q = -\frac{S(A)+S(A+1)}{2}.
	\label{chempot}
\end{align}

Let us consider the evolution of proton states. Fig.~\ref{Si_spe}(a,b) shows that for both parametrizations the calculations agree satisfactorily with the experimental data estimates \cite{Bes17} for the $1d_{5/2}$ and $2s_{1/2}$ states and underestimate higher states, while the SGII parametrization shows slightly better agreement in the region of neutron-rich isotopes. Inclusion of the tensor contribution does not affect the position of the $2s_{1/2}$ state, but affects the behavior of the $1d$ states and thus leads to some increase in the energy gap between $1d_{5/2}$ and $2s_{1/ 2}$, which, as we shall see later, affects the populations of the corresponding states. Changes in the position of the $1d$ and $1p$ states upon inclusion of the tensor component are consistent with Otsuka's rule. As $1d_{3/2}$ is filled with neutrons, the proton states $1d_{5/2}$ and $1p_{3/2}$ with $j_>$ attract more strongly, while the states $1d_{3/2}$ and $1p_{1/2}$ with $j_<$ are repelled. Further filling of the state $\nu 1f_{7/2}$ leads to the opposite effect. It can be seen that in the case of the SGII+T parametrization, the changes in the spin-orbit splitting of different levels are much more pronounced than in the case of SLy5+T. Apparently, in addition to the difference in the strength of the tensor interaction (in the SGII+T parametrization, the contribution of tensor forces is greater), the differences in the characteristics of the basic parametrizations also play a certain role. The response of the core structure to additional changes is stronger, with smaller values of $K_{\infty}$; in the case of SGII, the value of incompressibility is somewhat smaller. 

\end{multicols}
\begin{center}
\includegraphics[width=16cm]{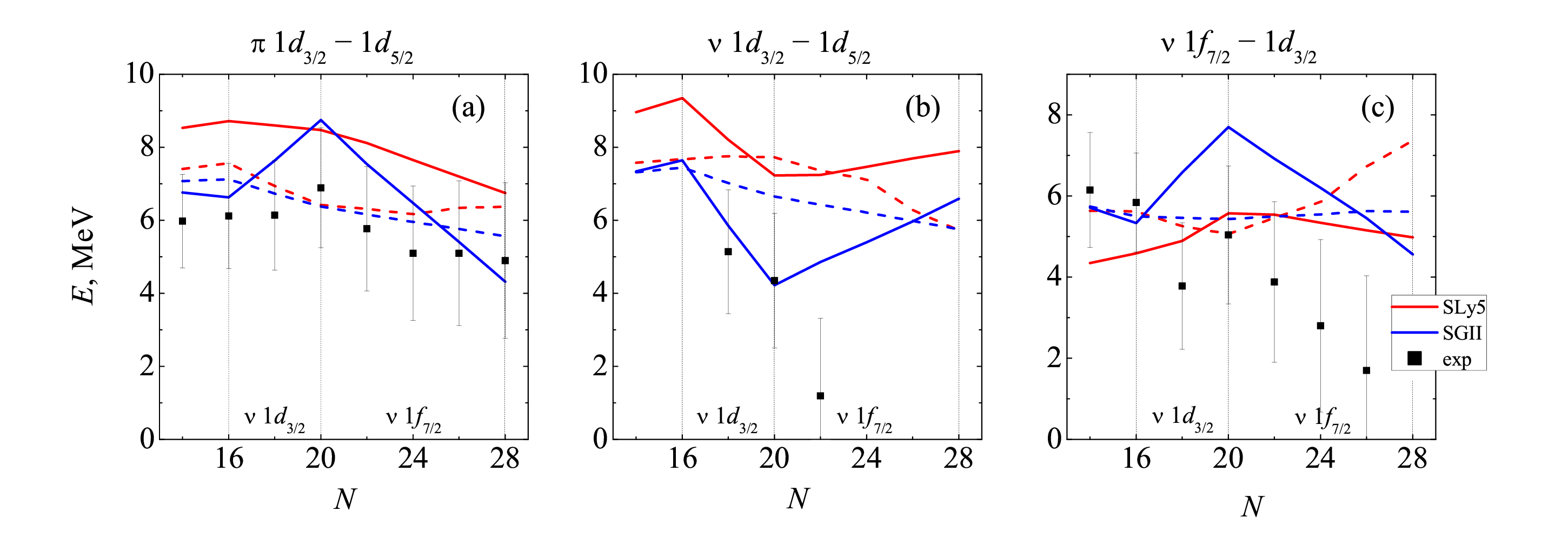}
\figcaption{\label{Si_diff} Splitting of proton (a) and neutron (b) states $1d_{3/2} - 1d_{5/2}$, as well as $1f_{7/2} - 1d_{3/2}$ states in silicon isotopes. Solid (dashed) lines show calculations with tensor forces (without taking tensor forces into account).}
\end{center}
\begin{multicols}{2}

Of great interest is the change in the spin-orbit splitting between proton $d$-states as the $1f_{7/2}$ shell is filled with neutrons. Experimental data show that the strongest splitting is present in the $^{34}$Si magic nucleus, however, quantitative estimates differ greatly and can reach 10 MeV. We have previously discussed this issue in detail in \cite{Bes17} and will use the estimates from that paper for comparison. Fig.~\ref{Si_diff}a shows the results of our calculations. It can be seen that both variants of interaction lead to a splitting of $d$-states in $^{34}$Si of about 9 MeV, and for other nuclei they also overestimate the said splitting. However, in this case, the SGII+T interaction succeeds at reproducing the behavior of the dependence at a qualitative level.

Regarding the influence of the neutron excess on the single-particle states of neutrons, here the presence of tensor interaction also leads to changes in SPEs, and the effect will be opposite compared to the case of $np$-interaction. As seen from Fig.~\ref{Si_spe}(c,d), as $\nu 1d_{3/2}$ is filled, the neutron states with $j_>$ are repelled, and with the filling $\nu 1f_{ 7/2}$ the attraction is strengthened by the shell. At the same time, the unsatisfactory description of experimental estimates for $1d2s$ states should be noted. On the other hand, it should be noted that the accuracy of the experimental data based on the spectroscopy of single-nucleon transfer reactions drops dramatically for the states lying much lower than the Fermi surface. Nevertheless, Fig.~\ref{Si_diff}b shows that for the range of nuclei up to $^{34}$Si, it is the inclusion of tensor forces that makes it possible to qualitatively reproduce the dependence of the neutron spin-orbit splitting of the $d$-states.

For neutron single-particle states, we can also consider the splitting between $1d_{3/2}$ and $1f_{7/2}$. The experimental data demonstrate a maximum in $^{34}$Si related to the manifestation of the magic number $N=20$ and a substantial drop to very small values as $\nu 1f_{7/2}$ is filled, which illustrates the decrease in the role of shell effects in highly neutron-rich silicon isotopes. As can be seen in Fig.~\ref{Si_diff}c, it is not possible to reproduce this character of the experimental dependence without including the tensor component, and the SGII+T parametrization makes it possible to reproduce the dependence at a qualitative level. The results with SLy5+T also give a maximum in $^{34} $Si, and the splitting value agrees with the experimental estimates.

It is important to note that the scale of the effect for neutrons is comparable to that observed earlier for protons (Fig.~\ref{Si_diff}a). In \cite{Otsuka05} it is shown that in the most elementary case, when $\alpha_T=\beta_T$, the interaction of like nucleons (governed purely by the isovector component of the forces) is 2 times weaker than $np$-interaction (dependant on both isoscalar and isovector components). Therefore, traditionally, when considering the effects associated with tensor interaction, one treats the influence of the neutron excess on the states of protons in isotopes or vice versa, the influence of the number of protons on the states of neutrons in isotones. In the case of interactions SLy5+T and SGII+T obtained by fitting the experimental data, $|\alpha_T|>|\beta_T|$, indicating that the contribution from the isovector component of tensor interaction may indeed be comparable to that from the $np$ tensor interaction. This inequality holds for quite a few of the existing interactions, including SLy4+T, SkP+T, SkO+T \cite{Zal08}, Sktxb \cite{Bro06} and more recent interactions SGII+T2 and SkO'+T \cite{Wu20}. All of these were obtained perturbatively with the tensor terms added on top of the fixed central part. It may also be of interest to check the case when a parameter set was generated via a variational procedure, with parameters of the central part refitted altogether as tensor forces are included. An example of such a set would be SAMi \cite{Roc12} with its recently obtained counterpart SAMi+T \cite{She19}. Notably, for SAMi+T $\alpha_T=-39.8$ and $\beta_T=66.7$ MeV fm$^5$, and the mentioned inequality does not hold. For SGII+T2, on the other hand, $\alpha_T=-162.5$ and $\beta_T=4.7$ MeV fm$^5$, making it a case where the isovector part responsible for interaction of like nucleons should be dominant.

\end{multicols}

\begin{center}
\tabcaption{ \label{tab2} Splitting between the proton and neutron $1d_{3/2}$ and $1d_{5/2}$ levels, as well as the neutron $1f_{7/2}$ and $1d_{3/2}$ obtained with various Skyrme force parametrizations in $^{28,34,42}$Si. Experimental values from \cite{Bes17} are shown at the end for comparison.}
\footnotesize
\begin{tabular*}{170mm}{@{\extracolsep{\fill}}c|ccc|ccc|ccc}
	\toprule & & $\pi 1d_{3/2}-1d_{5/2}$ & & & $\nu 1d_{3/2}-1d_{5/2}$ & & & $\nu 1f_{7/2}-1d_{3/2}$ & \\
	\hline
	Parametrization & $^{28}$Si & $^{34}$Si & $^{42}$Si & $^{28}$Si & $^{34}$Si & $^{42}$Si & $^{28}$Si & $^{34}$Si & $^{42}$Si \\
	\hline
	SGII & 7.08 & 6.38 & 5.60 & 7.31 & 6.65 & 5.98 & 5.74 & 5.43 & 5.61 \\
	SGII+T & 6.77 & 8.74 & 4.32 & 7.34 & 4.22 & 6.59 & 5.71 & 7.70 & 4.56 \\
	SGII+T2 & 8.91 & 8.16 & 7.02 & 9.35 & 7.13 & 8.31 & 3.81 & 5.00 & 3.18 \\
	\hline
	SLy5 & 7.41 & 6.42 & 6.37 & 7.58 & 7.73 & 5.73 & 5.63 & 5.07 & 7.36 \\
	SLy5+T & 8.54 & 8.48 & 6.75 & 8.96 & 7.23 & 7.90 & 4.35 & 5.58 & 4.98 \\
	\hline
	SLy4 & 7.60 & 7.03 & 6.27 & 7.87 & 7.30 & 6.47 & 5.36 & 5.51 & 6.55 \\
	SLy4+T & 8.60 & 7.65 & 7.35 & 8.90 & 8.06 & 7.51 & 4.40 & 4.78 & 5.53 \\
	\hline
	SAMi & 4.86 & 4.84 & 3.91 & 4.94 & 5.56 & 3.55 & 8.59 & 7.62 & 9.61 \\
	SAMi+T & 5.24 & 5.83 & 3.87 & 5.42 & 5.79 & 4.36 & 8.34 & 7.68 & 9.22 \\
	\hline
	exp & 6.0 $\pm$ 1.5 & 6.9 $\pm$ 1.6 & 4.9 $\pm$ 2.0 & & 4.2 $\pm$ 1.9 & & 5.8 $\pm$ 1.2 & 5.0 $\pm$ 1.5 & 1.8 $\pm$ 2.2 \\
	\bottomrule
\end{tabular*}%
\end{center}

\begin{multicols}{2}

For the sake of verifying the impact of different components of tensor forces on the single-particle structure of silicon isotopes, we performed some few additional calculations with interactions SLy4 and SLy4+T, SGII+T2, SAMi and SAMi+T in $^{28,34,42}$Si with filled $\nu 1d_{5/2}$, $sd$-shell and $\nu 1f_{7/2}$ respectively, and the results are shown in \ref{tab2}. We note the significant effect tensor forces have on the splitting between the neutron states $1d_{3/2,5/2}$ and $1f_{7/2}$ when SLy4+T and SGII+T2 forces are used. The $\nu 1f_{7/2} - 1d_{3/2}$ splitting in particular seems to be best reproduced within SGII+T2 Skyrme forces. The effect is indeed opposite in sign as compared to the Otsuka rules formulated for $np$-tensor interaction, for all the interactions under consideration. We note that SAMi predicts the opposite behaviour of the neutron level splittings, compared to the other interactions (maximum in $1d_{3/2,5/2}$ and minimum in $1f_{7/2} - 1d_{3/2}$ splitting in $^{34}$Si), but these extrema are less pronounced when tensor forces are taken into account. The largest impact on the proton states is observed with SGII+T and SAMi+T interactions, the latter describing the behaviour of the $\pi 1d_{3/2} - 1d_{5/2}$ splitting the best. 


For nickel isotopes, the general tendencies for the impact of the tensor force are the same, which brings about certain additional phenomena. Before discussing the results, we will first address the current state of available experimental data in neutron-rich nickel nuclei. In this work, we compare the calculated SPEs with those obtained from consistent stripping and pickup reaction analysis \cite{Bes11_1, Bes11_2, Bes09_1, Bes09_2, Sch13, Gr07, Vol07, Fla09, Smi04}. Analysis of stripping and pickup reactions is typically very sensitive to experimental conditions, resolution of the obtained spectra and the measurement range. For example, the authors of \cite{Bes09_2} compared the estimated SPEs of stable even $^{58-64}$Ni based off single out states assigned the largest spectroscopic factors, with the results taking into account the state fragmentation. The largest deviations amounted to 2 MeV far from the Fermi level, which may be of importance to states with small SPEs. Furthermore, while such deviations may not be critical for localization of individual single-particle states, the estimates of the spin-orbital splitting may differ significantly on a qualitative level in various approaches.

\end{multicols}
\begin{center}
\includegraphics[width=16cm]{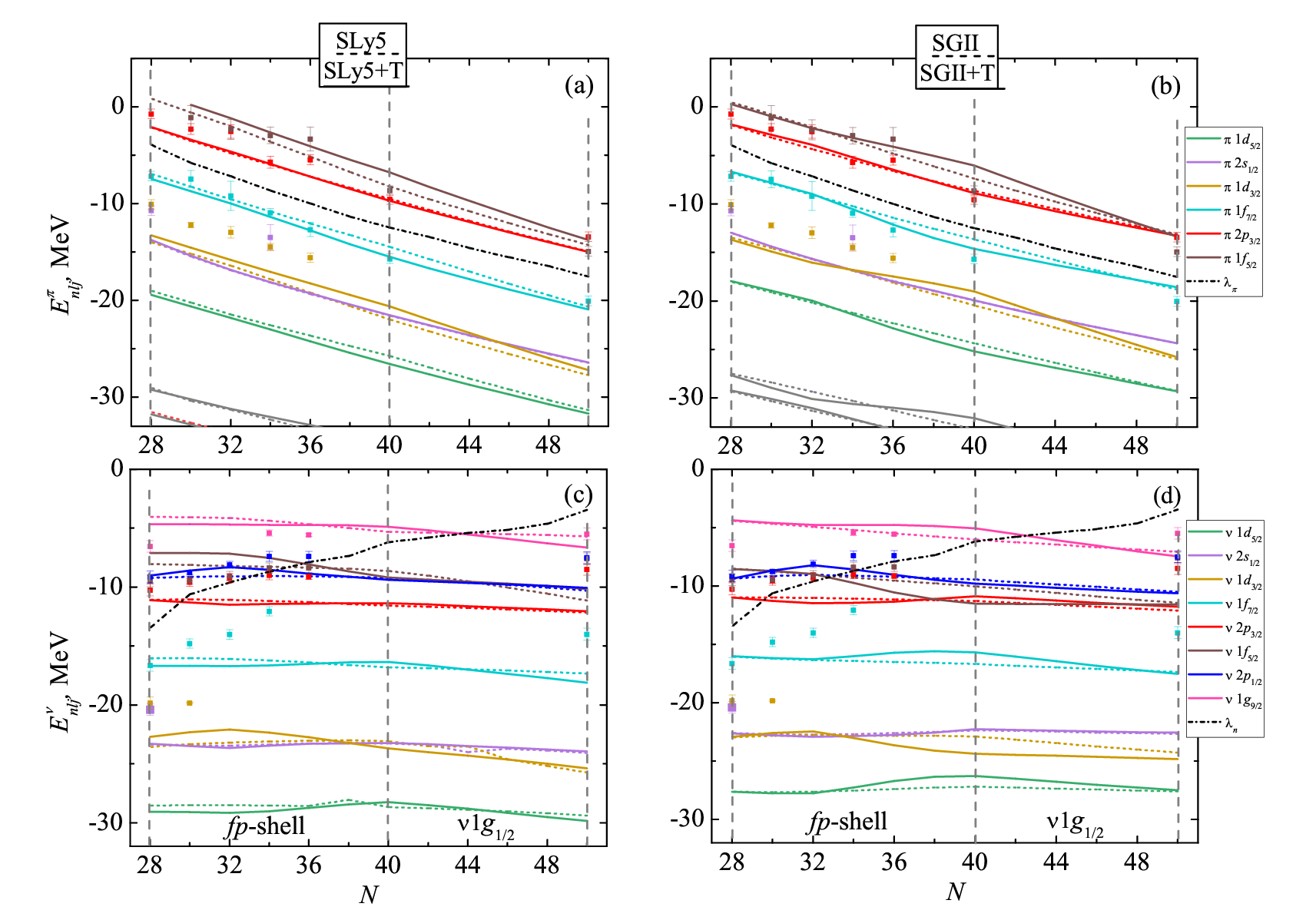}
\figcaption{\label{Ni_spe} SPEs in even nickel isotopes: proton (a, b) and neutron (c, d) states. Solid (dashed) lines show calculations with tensor forces (without taking tensor forces into account). The dashed-dotted line shows the chemical potential of protons (neutrons) in figures a and b (c and d). Experimental evaluated data are marked with dots. Experimental data for the proton states were taken from \cite{Gr07} for $^{56,78}$Ni and \cite{Bes11_1} for other isotopes. Experimental data for the neutron states were taken from \cite{Gr07} for $^{56,78}$Ni; $1f_{5/2}$ and $2p_{1/2, 3/2}$ states were taken from \cite{Sch13} for stable isotopes $^{58-64}$Ni, $1g_{9/2}$ and $1f_{7/2}$ states in these isotopes were taken from \cite{Bes11_1} and \cite{Vol07}, respectively.}
\end{center}
\begin{multicols}{2}

Our calculations of SPEs in neutron-rich nickel isotopes, together with experimental data from the sources listed earlier, are shown on Fig.\ref{Ni_spe}. For protons, of note is the inversion of states $1d_{3/2}$ and $2s_{1/2}$ taking place when tensor forces are accounted for. Their initial reversal as the neutron $fp$-shell is filled, and second reversal to the original order as $\nu 1g_{9/2}$ is filled, was predicted in \cite{Nak13} and recently in \cite{No21}, and is reproduced here using the SGII+T parametrization. In light of recently confirmed experimental data in copper isotopes \cite{Oli17} obtained at RIBF, of interest is also the behaviour of the splitting between the $\pi 2p_{3/2}$ and $\pi 1f_{5/2}$ states. The systematics of the first $3/2^-$ and $5/2^-$ states in these data suggest that the reversal of the corresponding proton orbitals takes place at around $N=45$. While this is achieved neither with the SLy5+T, nor with the SGII+T forces, SGII+T interaction does succeed at changing the ordering at $N=50$. A thorough comparison of this and other approaches in reproduction of the crossing between these levels in neutron-rich nickel isotopes, is also given in \cite{Ban22}.

When it comes to neutron states in neutron-rich nickel isotopes, neutrons initially fill the $fp$-shell, although the order of the filling within the shell appears to depend on the choice of the central part of interaction, as seen from Fig.~\ref{Ni_spe}. Furthermore introduction of the strong tensor component, such as in SGII+T, may also affect the order of certain levels. We note the influence of tensor forces on the neutron $1f_{5/2}$ state that gets additionally attracted as the $fp$-shell is filled, which results in it getting pushed below the $\nu 2p_{1/2}$ and then the $\nu 2p_{3/2}$ state. Experimental data taken from \cite{Bes17} show that $\nu 2p_{1/2}$ should indeed lie above the other states of the $fp$-shell, and SGII family interactions describe such ordering the best.

\end{multicols}
\begin{center}
\includegraphics[width=16cm]{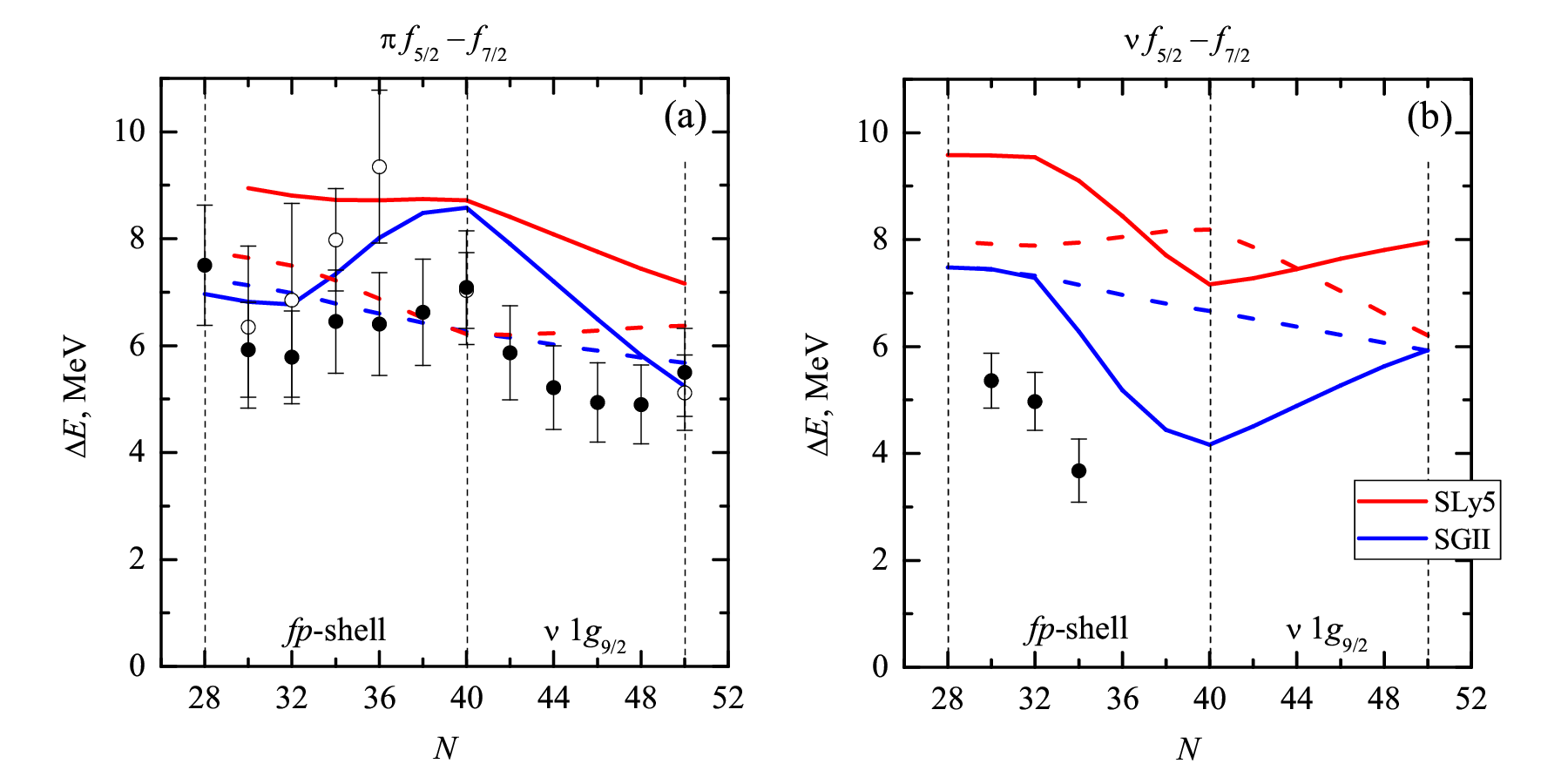}
\figcaption{\label{Ni_diff} Splitting of proton (a) and neutron (b) states $1f_{5/2} - 1f_{7/2}$ in nickel isotopes. Solid (dashed) lines show calculations with tensor forces (without taking tensor forces into account). For protons, experimental data from \cite{Bes11_1} and \cite{Gr07,Vol07,Fla09,Smi04,ENSDF} are shown with black and white circles, respectively. For neutrons, experimental data from \cite{Sch13} and \cite{Vol07} are shown with black circles.}
\end{center}
\begin{multicols}{2}

The tensor effects in nickel isotopes can be most clearly observed on the example of the magic gap between the proton or neutron $1f_{7/2}$ and $1f_{5/2}$ states (see Fig. \ref{Ni_diff}a and b, respectively). Various experimental data for these splittings appear to differ by up to 2-3 MeV for various nuclides, but the general pattern described by the Otsuka rule emerges clearly. We note that inclusion of tensor forces allows for description of the local maximum and minimum in protons and neutron level splittings, respectively. The general scale of the effect is, once again, better described by the interaction SGII+T. Results of additional calculations performed with interactions SLy4(+T), SGII(+T2) and SAMi(+T) are presented in Table~\ref{tab3}. While SAMi+T with the stronger $np$-component does reproduce qualitatively the maximum of the $\pi 1f_{7/2,5/2}$ splitting for $^{68}$Ni, it also severely underestimates said splitting on the entire chain of nickel isotopes, while SGII and SLy4 appear to give more reasonable values. Although no experimental data is currently available for the difference between the neutron $\nu 1f_{7/2,5/2}$ SPEs in $^{56,68,78}$Ni, some few data in $^{30,32,34}$Ni indicate the decrease of the splitting towards $^{68}$Ni (as seen in Fig.~\ref{Ni_diff}), which also is reproduced with SGII+T2 and SLy4+T. We suspect this suggests the correct expectation that the isovector part of tensor forces should be prominent, and that the Otsuka rule may indeed work in the opposite way for $np$ tensor forces and tensor forces between like nucleons.

\end{multicols}

\begin{center}
\tabcaption{ \label{tab3} Splitting between the proton and neutron $1f_{5/2}$ and $1f_{7/2}$ levels obtained with various Skyrme force parametrizations in $^{56,68,78}$Si. Experimental values from \cite{Bes11_1, Gr07} are shown at the end for comparison.}
\footnotesize
\begin{tabular*}{170mm}{@{\extracolsep{\fill}}c|ccc|ccc}
	\toprule & & $\pi 1f_{5/2}-1f_{7/2}$ & & & $\nu 1f_{5/2}-1f_{7/2}$ & \\
	\hline
	Parametrization & $^{56}$Ni & $^{68}$Ni & $^{78}$Ni & $^{56}$Ni & $^{68}$Ni & $^{78}$Ni \\
	\hline
	SGII & 7.26 & 6.27 & 5.67 & 7.52 & 6.67 & 5.92 \\
	SGII+T & 6.97 & 8.58 & 5.24 & 7.48 & 4.16 & 5.94 \\
	SGII+T2 & & 8.16 & 7.21 & 9.74 & 6.82 & 7.98 \\
	\hline
	SLy5 & 7.77 & 6.22 & 6.37 & 7.97 & 8.19 & 6.21 \\
	SLy5+T &  & 8.72 & 7.16 & 9.58 & 7.16 & 7.95 \\
	\hline
	SLy4 & 8.07 & 7.08 & 6.47 & 8.37 & 7.53 & 6.77 \\
	SLy4+T & & 7.65 & 7.51 & 9.57 & 8.22 & 7.79 \\
	\hline
	SAMi & 4.42 & 4.54 & 3.54 & 4.44 & 5.70 & 3.26 \\
	SAMi+T & 4.75 & 5.90 & 3.60 & 4.92 & 5.80 & 3.85 \\
	\hline
	exp & 7.5 $\pm$ 1.6 & 7.0 $\pm$ 1.0 & 5.5 $\pm$ 0.9 & & & \\
	\bottomrule
\end{tabular*}%
\end{center}

\begin{multicols}{2}
	
It should be pointed out that the tensor terms in SAMi+T in particular were fitted using the pseudodata of the neutron-proton drops coming from  relativistic Brueckner-Hartree-Fock (RBHF) calculations, rather than to experimental data. The approach of comparing to meta-data generated within \textit{ab initio} calculations allows for narrowing down the constraints on the possible nuclear density functionals, and has the benefit of avoiding the particle-vibration coupling which typically makes it harder to analyse experimental data. In recent works \cite{Wan21,Zha20,She18} employing the RBHF approach, it was shown that tensor forces in neutron and neutron-proton drops may work similarly for proton and neutron states as neutrons are added in the system. This result was indeed also obtained in \cite{She19} where interaction SAMi+T was originally proposed. On the example of \ref{tab2} and \ref{tab3} it is clearly seen that SAMi+T often predicts maxima for neutron level splittings in cases when minima are obtained with other forces, and vice versa. The reasons can be traced down to the central part of the interaction, however, as the same behaviour is obtained with SAMi without tensor terms. Evidently, these extrema in $^{34}$Si and $^{68}$Ni are smoothed out for splittings between neutron states, and become more prominent for splittings between proton states as these terms are taken into account, meaning that even here the isovector and $np$ tensor components work in the opposite directions.

Let us consider the effect of tensor forces on the occupation of single-particle states near the Fermi level. In the BCS approach, this characteristic depends on the energy of the single-particle state $E_{nlj}$:
\begin{align}
	n_{nlj} = \frac{1}{2} \left( 1 - \frac{E_{nlj}-\lambda}{\sqrt{(E_{nlj}-\lambda)^2 + \Delta^2}} \right),
\end{align}
where $\lambda$ is the chemical potential and $\Delta$ is the energy gap. Since accounting for tensor interaction results in a change in the energies of single-particle states, these changes should be reflected in the corresponding occupation numbers $n_{nlj}$.

\end{multicols}
\begin{center}
\includegraphics[width=16cm]{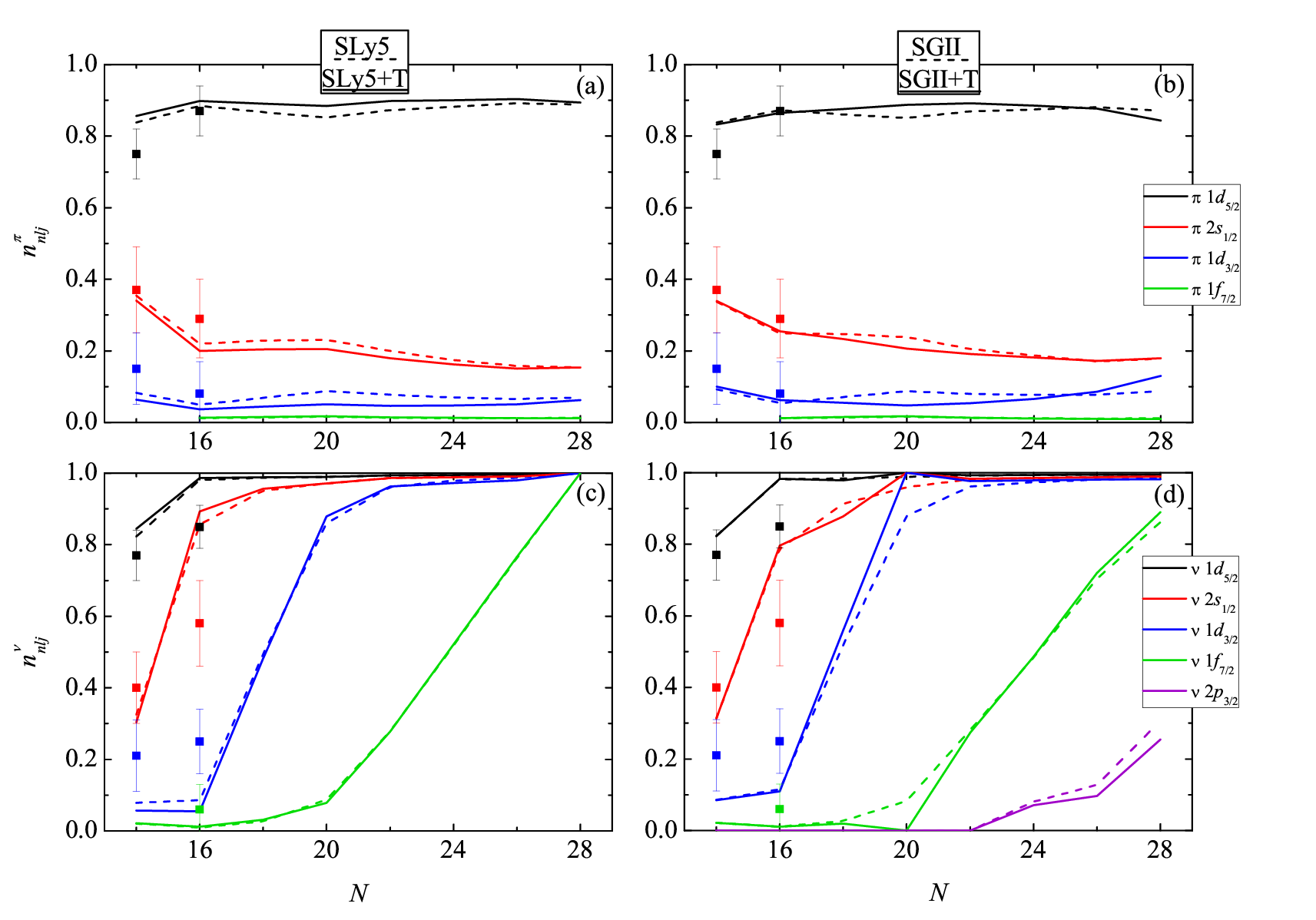}
\figcaption{\label{Si_occ} Occupation numbers of (a, b) proton and (c, d) neutron single-particle levels near the Fermi surface in silicon isotopes. Solid (dashed) lines show calculations with tensor forces (without taking tensor forces into account). Experimental data \cite{Bes17} are marked with dots.}
\includegraphics[width=16cm]{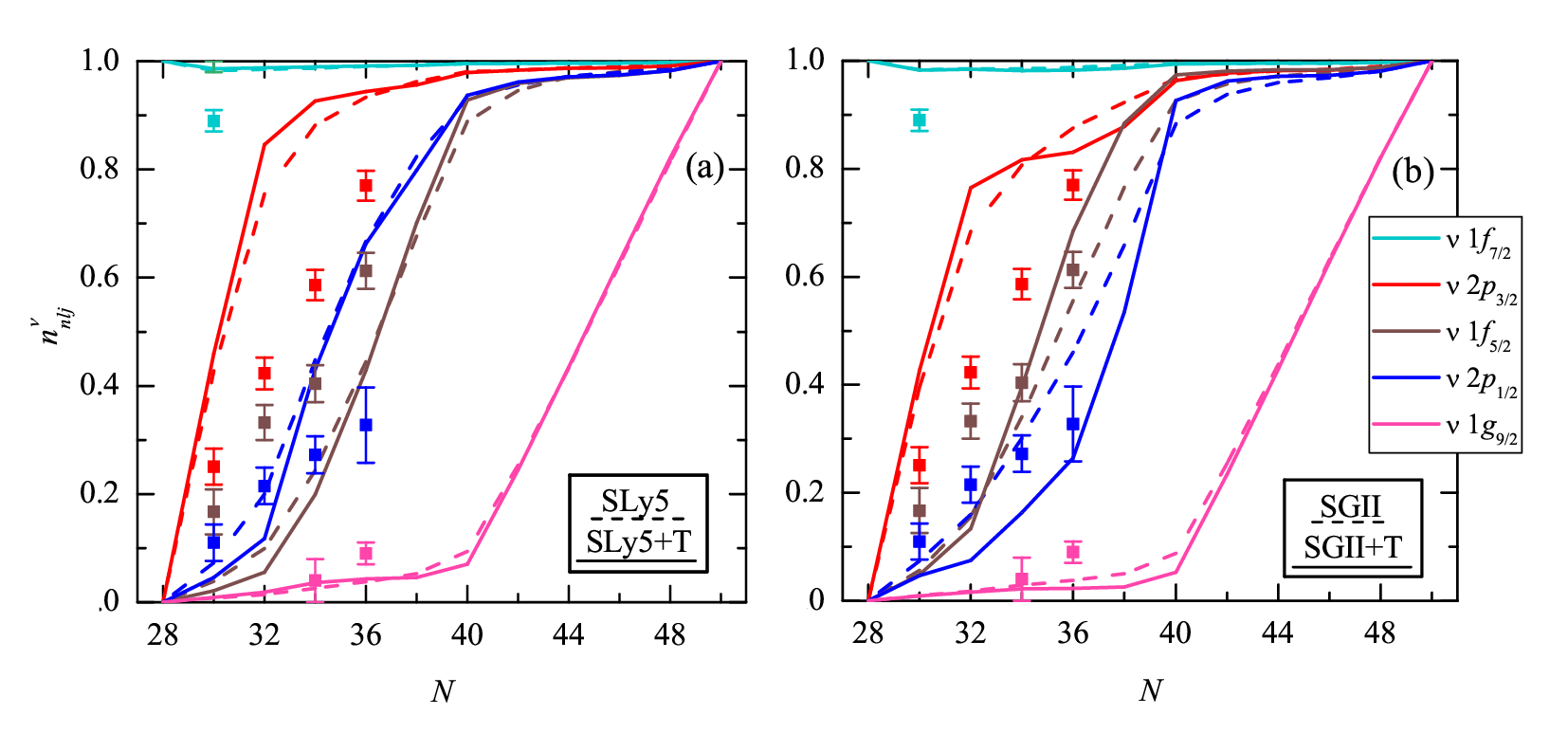}
\figcaption{\label{Ni_occ} Occupation numbers of neutron single-particle levels near the Fermi surface in nickel isotopes. Solid (dashed) lines show calculations with tensor forces (without taking tensor forces into account). Experimental data \cite{Sch13} for the $1f_{5/2}$ and $2p_{1/2, 3/2}$ states and \cite{Bes11_1} for state $1g_{9/2}$ are marked with dots.}
\end{center}
\begin{multicols}{2}

The calculated single-particle occupation numbers for even silicon isotopes are presented on Fig.~\ref{Si_occ}. Both parametrizations, SLy5+T and SGII+T, effectively reduce the effects of nucleon pairing. Pairing correlations of nucleons lead to smearing of the Fermi level, with states above $\lambda_q$ getting partially filled. Accounting for the tensor interaction leads to a decrease in this effect: the population of the levels with $E_{nlj}<\lambda$ (see~Fig.~\ref{Si_spe}) increases. This is most clearly manifested in calculations for the $^{34}$Si isotope, where in the case of the SGII+T parametrization with a large tensor component, neutron pairing actually disappears. Using proton states as an example, one can see that in silicon isotopes the Fermi level lies between the $\pi 1d_{5/2}$ and $\pi 1 d_{3/2}$ levels. When the contribution of tensor forces is taken into account, the splitting between these states increases; accordingly, as a result of moving away from the Fermi level, the population of subshells with energies above $\lambda_p$ decreases. The population of neutron states changes according to the same patterns: an increase in the splitting between the $\nu 1f_{7/2}$ and $\nu 1d_{3/2}$ levels, between which the Fermi level passes for nuclei near $^{34}$Si, effectively leads to a decrease in pairing correlations. It is important to note that the results also strongly depend on the properties of the central part of the interaction: in the case of SLy5, the neutron level $2p_{3/2}$ can turn out to be both in the continuum for all neutron-rich isotopes (Fig.~\ref{Si_occ}c), and in the case SGII in $^{38-42}$Si isotopes, it is in the potential well (Fig.~\ref{Si_occ}d). Thus, in the case of SLy5(+T), the magic core $^{42}$Si is closed. As in the case of SPEs, theoretical calculations do not agree with experimental estimates of the populations of neutron states, which suggests that, in addition to tensor forces, other effects must be taken into account, primarily the deformation of nuclides.

For nickel isotopes, we only considered neutron pairing, as $Z=28$ is a magic number. Tensor forces result in much the same phenomenon: nucleons are pushed below the Fermi level, effectively decreasing the pairing correlations. Here, once again, we notice the difference in occupation order between neutron states $2p_{1/2}$ and $1f_{5/2}$. This correlates with the order of these states on Fig.~\ref{Ni_spe}, and the correct reproduction of ordering from experimental data is achieved with SGII interactions. Admittedly, SGII without the tensor component appears to give better results here, which comes as a result of the inconsistencies in the fitting procedure.

\section{Conclusion}

The structure of even silicon isotopes $^{28-42}$Si and nickel isotopes $^{56-78}$Ni was calculated within the framework of the self-consistent Hartree-Fock approach with the Skyrme interaction. Our task was to analyze the influence of the tensor interaction component on changes in single-particle characteristics with an increase in the excess of neutrons by the example of the SLy5+T and SGII+T parametrizations, which include tensor forces of various magnitudes. When Skyrme forces are used, tensor forces contribute to the so-called $J^2$ terms, and thus their influence leads to a change in the spin-orbit splitting of single-particle levels.

Both tensor forces and pair correlations lead to an increase in the binding energy. As a consequence, taking into account these effects leads to overestimation of the specific binding energy of all considered silicon and nickel isotopes, and the best agreement is obtained for the parametrization of SLy5 in the absence of additional effects. In this case, the SGII+T interaction leads to a particularly strong overbinding of nuclei, which is associated with the peculiarities of SGII fitting protocol. Thus, it should be recognized that inclusion of the tensor contribution on top of the parametrizations previously fitted to various experimental data, strictly speaking, can only be used for test calculations. For a more accurate quantitative reproduction of the experimental estimates of the energy characteristics, it is necessary to fit the Skyrme parameters via a variational procedure affecting both the tensor and central parts.

Accounting for the tensor forces, however, improves the description of the splitting of various single-particle levels in neutron-rich even silicon and nickel isotopes. Changes in the energies of proton single-particle states in silicon isotopes, and, accordingly, the dependence of the spin-orbit splitting of the $d$-states with an increase in the number of neutrons, are in full accordance with Otsuka rules: filling the $j'_<$ level with neutrons leads to an increase in the spin-orbit splitting between proton levels, and when $j'_>$ is filled, this splitting, on the contrary, decreases. The results of calculations for neutron states show that the changes in the spin-orbit splitting of neutron levels are reversed, but the magnitude of the change is comparable in absolute value. In this case, the dependence of the splitting of $d_{3/2}-d_{5/2}$ levels, as well as $sd$- and $f$-shells, on the number of neutrons in silicon isotopes is qualitatively better reproduced with the SGII+T forces. For nickel, tensor forces additionally induce several changes in the ordering of various single particle states, namely, neutron and proton $2s_{1/2}$ and $1d_{3/2}$, as well as the states of the neutron $fp$-shell. The general behaviour of the splitting between the proton and neutron $f_{5/2} - f_{7/2}$ states was reproduced with SGII+T interaction, although the SHF approach appears to be insufficient when it comes to description of the energies of individual states.

According to spectroscopic data in neutron-rich silicon and nickel isotopes, the splitting between various neutron single-particle states appears to change in the opposite way compared to the proton states, along the isotopic chains. This behaviour of neutron state splitting appears to be best reproduced when interactions SGII+T, SGII+T2, SLy5+T and SLy4+T are employed. All of these parametrizations are indeed characterised by the $np$ and isovector tensor terms of opposite signs and comparable amplitudes. Parametrization SAMi+T, on the other side, predicts similar behaviour of the protons and neutron state splittings, which is attributed to the peculiarities of its central part. While the contribution from the isovector component of tensor forces is noticeably smaller compared to $np$ tensor forces, these two terms still act in the opposite way, in agreement with other interactions.

Analysis of the populations of both proton and neutron levels near the Fermi surface showed that the inclusion of tensor forces effectively leads to a decrease in pairing in silicon and nickel isotopes. This effect is related to the fact that the tensor interaction additionally pushes the levels closest to $\lambda_{n,p}$ away from the Fermi surface. Such filling is specific for silicon and nickel isotopes, and in other nuclei one can apparently expect both similar and opposite manifestations of tensor forces.

\end{multicols}

\end{CJK*}
\end{document}